\documentclass[manuscript,screen]{acmart}

\usepackage[export]{adjustbox}
\usepackage{graphicx}
\usepackage{balance}  
\usepackage{multirow}
\usepackage{hyperref}
\usepackage{thumbpdf}
\usepackage{booktabs}
\usepackage{xcolor}
\usepackage{times}
\usepackage{url}

\usepackage{graphicx}
\usepackage{subfigure}
\usepackage{xfrac}
\usepackage{fix-cm}
\usepackage{array}
\usepackage[normalem]{ulem}
\usepackage{xspace}
\usepackage{wrapfig}
\usepackage{tcolorbox}
\usepackage{enumitem}

\newcommand\name{\textit{I\&D}\xspace}

\def\ie{\textit{i.e.},~}
\def\etal{\textit{et al.}~}

\def\eg{\textit{e.g.},~}

\def\Snospace~{\S{}}

\setcopyright{acmcopyright}

\newcommand{\point}[1]{\vspace{1mm}\noindent\textbf{#1}.}

\newcommand{\changed}[1]{\textcolor{black}{#1}}

\newcommand{\insight}[1]{
\begin{center}
\begin{tcolorbox}[colback=gray!10,
                  colframe=gray!10,
                  arc=2mm, auto outer arc,
                  boxrule=0.5pt,
                 ]{#1}
\end{tcolorbox}
\end{center}} 

\begin{document}


\title{Enhanced Fairness Testing via Generating Effective Initial \\ Individual Discriminatory Instances}


\author{Minghua Ma, Zhao Tian}
\email{minghuama@microsoft.com}
\email{v-zhaotian@microsoft.com}
\affiliation{%
  \institution{Microsoft Research Lab Asia}
  \city{Beijing}
  \country{China}
  \postcode{100080}}

\author{Max Hort, Federica Sarro}
\email{max.hort.19@ucl.ac.uk}
\email{f.sarro@ucl.ac.uk}
\affiliation{%
  \institution{University College London}
  \city{London}
  \country{UK}}

\author{Hongyu Zhang}
\email{hongyu.zhang@newcastle.edu.au}
\affiliation{%
  \institution{The University of Newcastle}
  \city{Callaghan, NSW}
  \country{Australia}}

\author{Qingwei Lin, Dongmei Zhang}
\email{qlin@microsoft.com}
\email{dongmeiz@microsoft.com}
\affiliation{%
  \institution{Microsoft Research Lab Asia}
  \city{Beijing}
  \country{China}
  \postcode{100080}}

\begin{abstract}
Fairness testing aims at mitigating unintended discrimination in the decision-making process of data-driven AI systems.

Individual discrimination may occur when an AI model makes different decisions for two distinct individuals who are distinguishable solely according to protected attributes, such as age and race. Such instances reveal biased AI behaviour, and are called Individual Discriminatory Instances (IDIs).

In this paper, we propose an approach for the selection of the initial seeds to generate IDIs for fairness testing.
Previous studies mainly used random initial seeds to this end. However this phase is crucial, as these seeds are the basis of the follow-up IDIs generation. We dubbed our proposed seed selection approach \textit{I\&D}. It generates a large number of initial IDIs exhibiting a great diversity, aiming at improving the overall performance of fairness testing.

Our empirical study reveal that \textit{I\&D} is able to produce a larger number of IDIs with respect to four state-of-the-art seed generation approaches, generating 1.68X more IDIs on average. Moreover, we compare the use of \textit{I\&D} to train  machine learning models and find that using \textit{I\&D} reduces the number of remaining IDIs by 29\% when compared to the state-of-the-art, thus indicating that \textit{I\&D} is effective for improving model fairness.

\end{abstract}

\ccsdesc[500]{Software and its engineering~Software verification and validation}
\ccsdesc[500]{Computing methodologies~Machine learning}


\keywords{Fairness testing, machine learning models, software fairness}

%



\maketitle

\section{Introduction}
\label{sec:intro}
Artificial Intelligence (AI) systems have become increasingly popular over the years, and are now at the core of many data-driven software systems such as loan approval and risk assessments \cite{wang2020deep}.
Although machine learning models have achieved significant performance improvements, their fairness remains a prominent concern that needs to be addressed~\cite{chakraborty2021bias, biswas2020machine, pham2020problems}. 

Software fairness testing sets out to reveal fairness bugs of software system (\ie situations that reveal bias)~\cite{chen2022fairness}.
Among fairness testing goals, \textit{individual discrimination} has been addressed frequently~\cite{angell2018themis,dwork2012fairness}.
Individual discrimination occurs when a machine learning model yields different results for the instances that can only be distinguished by one or more protected attributes, such as age, race, or gender \cite{angell2018themis, imtiaz2019investigating}.
For example, two individuals with different gender (\eg female and male), all other attributes being equal, should receive the same response when applying for a bank loan.

\begin{figure}[]
    \centering
    \subfigure[A pair of IDIs]{
		\includegraphics[width=.3\linewidth]{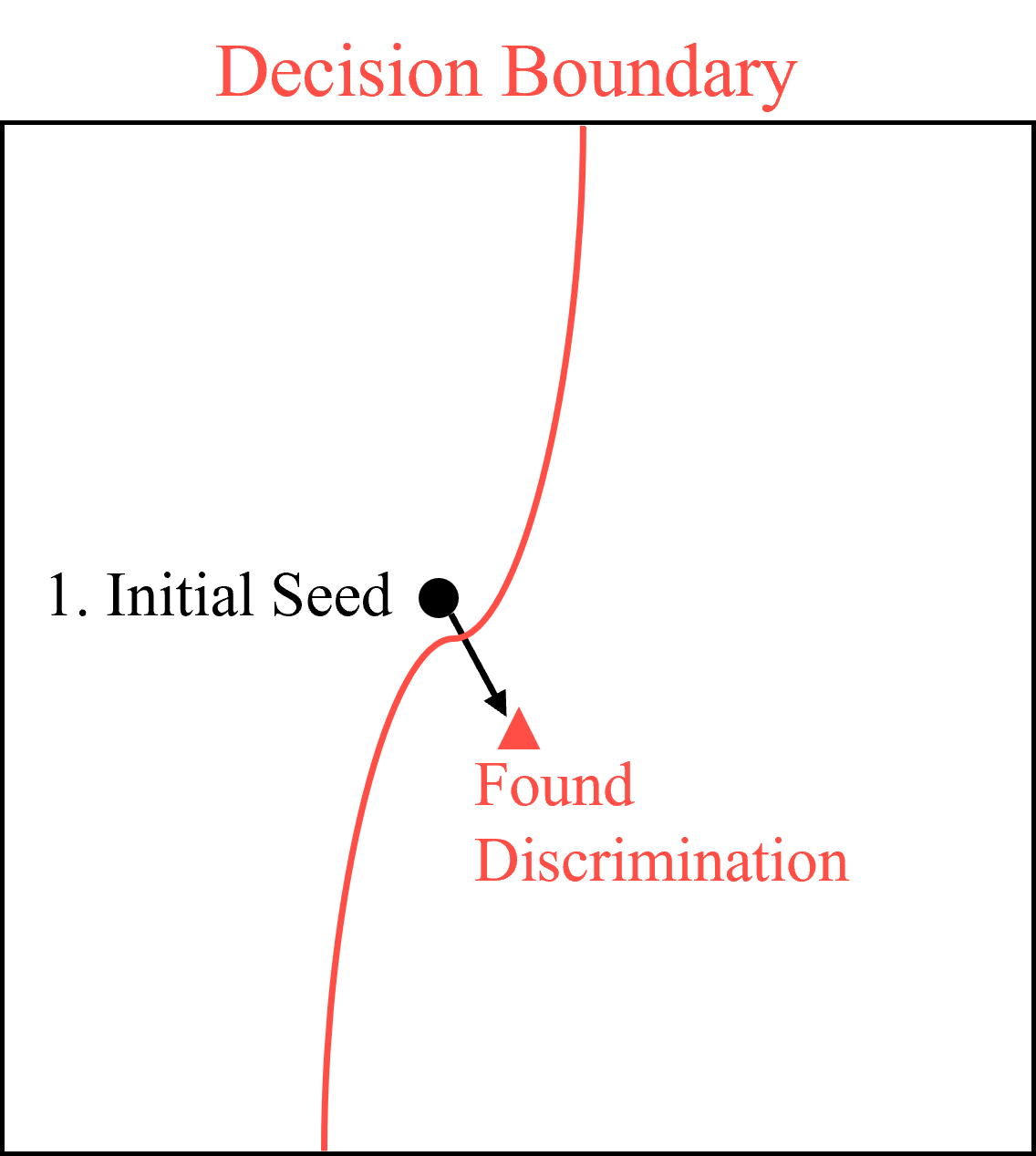}
		\label{fig:demo1}
    }
	\subfigure[Generation]{
		\includegraphics[width=.3\linewidth]{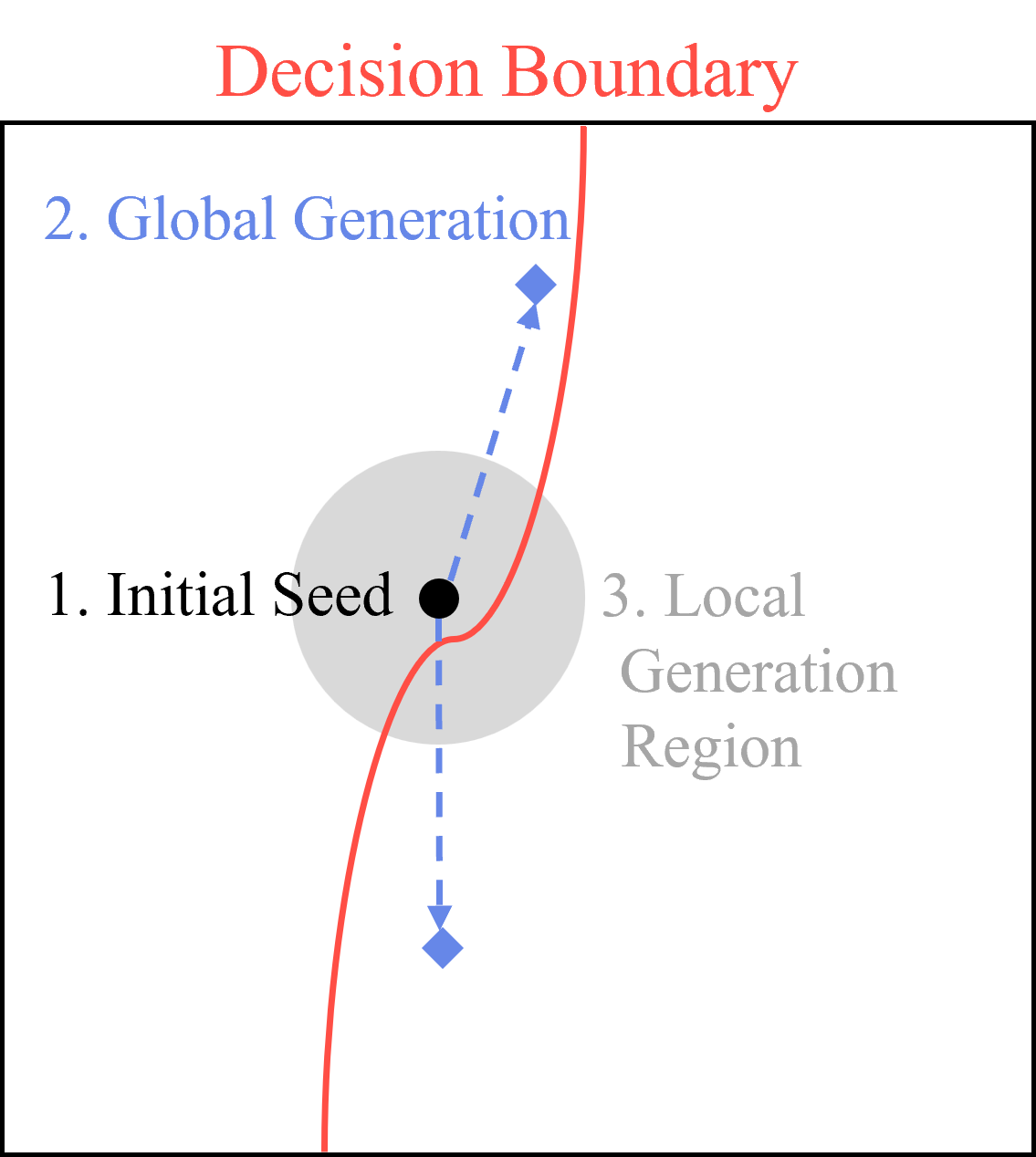}
		\label{fig:demo2}
	}
    \caption{Sketch map of individual discriminatory instances (IDIs) and their generation process.}
    \label{fig:demo}
\end{figure}

To date, fairness testing of machine learning and AI systems has become a topic of interested in Software Engineering research~\cite{chen2022fairness, SoremekunFairnessSurvey}.
This includes approaches to generating individual discriminatory instances (IDIs) for fairness testing~\cite{chen2022fairness}, such as AEQUITAS \cite{udeshi2018automated}, Symbolic Generation (SG) \cite{aggarwal2019black},
ADF \cite{zhang2020white}, and EIDIG \cite{zhang2021efficient}.
All these existing approaches follow the same three-phase framework, as depicted in \autoref{fig:demo2}.
First, they choose \textit{initial seeds} from a given dataset, \ie the black dot in \autoref{fig:demo2}, according to a given strategy.\footnote{Specifically, AEQUITAS selects seeds from the input data using a random sampling mechanism, while other approaches, such as SG, ADF, and EIDIG, cluster the input data using K-means, and then choose initial seeds from each cluster on a round-robin basis.}
Second, they perform a \textit{global generation} to explore a wider range of IDIs, which is illustrated by the blue squares in \autoref{fig:demo2}. 
Once an IDI is found during the \textit{global generation} phase, a \textit{local generation} is performed in the third step.
The \textit{local generation} searches for further IDIs in the neighborhood of the collected IDIs from the \textit{global generation} phase, which is depicted in gray in \autoref{fig:demo2}. 
Afterwards, the model can be retrained to minimize discrimination using the generated IDIs.

The existing approaches focus on improving the latter two phases to boost the overall performance of fairness testing.
For example, AEQUITAS randomly perturbs attribute values in the local generation phase.
SG globally creates a decision tree and then performs symbolic generation for the local generation.
On neural network models, ADF and EIDIG employ an adversarial sampling technique during both the global and local generation phases.
Regarding the first phase (\ie initial seeds selection), existing approaches adopt random or clustering-based sampling strategies.
However, we argue that this phase is very important to the overall performance of fairness testing.
Specifically, since this phase is the basis of the follow-up global and local generation phases, the quality of initial seeds could directly affect the performance of IDI generation.
On the one hand, 
as demonstrated in the existing study~\cite{zhang2020white, zhang2021efficient}, the number of generated IDIs in the two-generation phases grows linearly with the number of IDIs in the initial seeds.
That is, the more initial IDIs provided, the more IDIs may be generated, which is advantageous for fairness testing.
On the other hand, initial IDIs with a high diversity could facilitate generating a larger variety of IDIs, which can improve the retraining and debiasing of machine learning models.
That is, both the number of IDIs and the diversity of IDIs are pursued in practice, which can be controlled by the quality of the initial IDIs, and thus designing a more effective initial seed selection strategy is desired.


In this work, we propose a novel initial seed selection approach, named \name. 
Our approach aims to obtain more, diverse initial IDIs, and thus improve the overall performance of fairness testing.
That is, the contribution of \name is orthogonal to existing fairness testing approaches.
\name can be integrated with any existing fairness testing approaches by replacing their initial seed selection method with \name, in order to further improve their performance.

\name is required to address two major challenges:
1) How do we effectively select seeds to obtain more initial IDIs?
2) How do we improve the diversity of initial IDIs?
For the first challenge, we design a novel IDI initialization algorithm that generates a ``chiral'' model,\footnote{Chirality is a feature of asymmetry that is important in many research works \cite{cahn1966specification}. An object (model) is chiral if it can be distinguished from its mirror image (trained by mutated data); that is, it cannot be superimposed onto it.} which is trained by mutating the protected attributes of the training data.
It is more likely to identify initial IDIs if the chiral model predicts differently from the original model, for the same instance.
To overcome the second challenge, we propose to use the SHAP value \cite{lundberg2017unified}, which is a game-theory-based approach for explaining the output of any machine learning model, to explain the difference in prediction behavior between the chiral and the original model for each initial IDI.
Then, \name clusters the initial IDIs based on their
SHAP values, and selects diverse initial IDIs from each cluster in a round-robin way for future usage in the global and local generation phases.

To evaluate the effectiveness of \name, we undertake a thorough empirical evaluation using three open-source datasets that have been widely used in prior studies~\cite{udeshi2018automated, aggarwal2019black, zhang2020white, zhang2021efficient, chakraborty2020fairway}.
To investigate whether \name boosts the performance of existing IDI generation approaches, such as AEQUITAS, SG, ADF, and EIDIG, we integrate \name with each of them by replacing their initial seed selection strategy with \name, and investigate four different Machine Learners (namely Logistic Regression, Support Vector Machines, Decision Trees, and Multi-layer Perceptron Classifier).

The empirical results reveal that \name can effectively obtain improved initial seeds, and significantly outperform all the compared approaches with their original initial seeds.
For example, the average number of generated IDIs with \name is 3038, while the value achieved by original approaches is just 1803. The improvement is 1.68X.
The results show that this is a promising approach for improving IDI initialization. 
Our results also show that using \name can improve the model fairness of the ML models considered herein, when such a model is re-trained with the IDIs generated by \name.

Overall, the key contributions of our work are as follows:
\begin{itemize}
\item To the best of our knowledge, this is the first study dedicated to the selection of initial IDIs for improving fairness testing.

\item To produce more, diverse initial IDIs, we design and implement a method called \name through building a chiral model and measuring the SHAP value of each initial IDI for explaining the prediction difference between the chiral and original models.

\item We conduct an extensive empirical assessment of \name based on three public datasets 
and four existing IDI generation approaches. The results show that \name can effectively boost the performance of these approaches in terms of both the number of generated IDIs and the reduced discrimination after retraining with the IDIs found. 

\end{itemize}


The replication package for this paper, including all our data,
source code, and documentation, is publicly available online at:

\centerline{\textbf{\url{https://anonymous.4open.science/r/fairness-095F/}}}

\section{Background}
\label{sec:background}
In this section, we first introduce the concept and notation of individual discrimination. 
Then, we review the existing IDI generation approaches.
Finally, we describe the SHAP value, which is a commonly adopted model explanation approach.

\subsection{Individual Discrimination}
\label{sec:bg_individual}
AI systems utilise various types of machine learning models, including decision trees \cite{loh2011classification}, regression analysis \cite{weisberg2005applied}, and neural networks \cite{sarle1994neural}.
Following prior studies on fairness testing \cite{udeshi2018automated, aggarwal2019black, zhang2020white, zhang2021efficient}, we focus on the binary classification problem, which is important in AI systems \cite{loh2011classification}. 
To demonstrate the generality of our approach, we do not focus on a specific type of machine learning model in our study.
We denote the machine learning model $\mathcal{M}: X \rightarrow Y$, and it generates a predicted class label $y \in Y$ with the highest probability for a given instance $x \in X$. 
$A = {A_1, A_2, ..., A_n}$ represents a set of attributes (features) in $X$. 
Assuming that each attribute $A_i$ ($i \in [1, n]$) has a domain value space of $\mathbb{I}_i$, then the total input domain of $x$ is equal to all possible combinations of attribute value spaces, \ie $\mathbb{I} = \mathbb{I}_1 \times \mathbb{I}_2 \times \ldots \times \mathbb{I}_n$.

Finding and generating individual discriminatory instances (IDIs) for a given machine learning model is the first step towards reducing discrimination and achieving individual fairness \cite{udeshi2018automated, aggarwal2019black, zhang2020white, zhang2021efficient}.
Discrimination is frequently described in terms of a group of protected attributes, such as age, race, and gender.
Individual discrimination occurs when a machine learning model makes different decisions for two identical individuals apart from protected attributes.
Note that the list of protected attributes is often application-specific and unrelated to the prediction goal, which is provided in advance \cite{zhang2020white}.
Deleting the protected attributes from the training data would not eliminate the bias, since individual discrimination may remain due to various co-relations between protected and non-protected qualities \cite{chakraborty2020fairway}.

Formally, the IDI $x$ of a machine learning model can be defined as follows:
$$ 
\left\{
\begin{aligned}
\exists p \in P, x_p \neq x'_p \\
\forall q \in NP, x_q = x'_q \\
f(x) \neq f(x')
\end{aligned}
\right.
$$
where $x'$ exists in $\mathbb{I}$, $P \subset A$ is a set of protected attributes like race and gender. 
$NP \subset A$ is the set of non-protected attributes, $P \cup NP = A$, and $P\cap NP = \varnothing$. 

We use the Census Income dataset as a running example.\footnote{https://archive.ics.uci.edu/ml/datasets/adult}
\autoref{sec:settings} contains more information about this dataset.
From this dataset, we consider the following two instances $x$ and $x'$: 
$$x: [3,5,3,0,2,8,3,0,\textbf{1},2,0,40,0,0], \ f(x) = 0 $$
$$x': [3,5,3,0,2,8,3,0,\textbf{0},2,0,40,0,0], \ f(x') = 1 $$

In the list, the attributes of an instance are represented as integers that are the model's input.
Gender, which is displayed in bold, is assumed to be the protected attribute here.
$x$ denotes a male and $x'$ denotes a female.
Except for gender, we can see that $x$ and $x'$ have identical attribute values. 
Since the model $\mathcal{M}$ has different prediction outcomes $f(x)$ and $f(x')$, we say that $x$ and $x'$ are a pair of IDIs for the model. 

In summary, machine learning models could make biased decisions for IDIs, which is destructive to fairness. 
It is important to effectively generate IDIs and improve the fairness of the machine learning model through testing and retraining.



\begin{figure}[]
    \centering
    \includegraphics[width=.\linewidth]{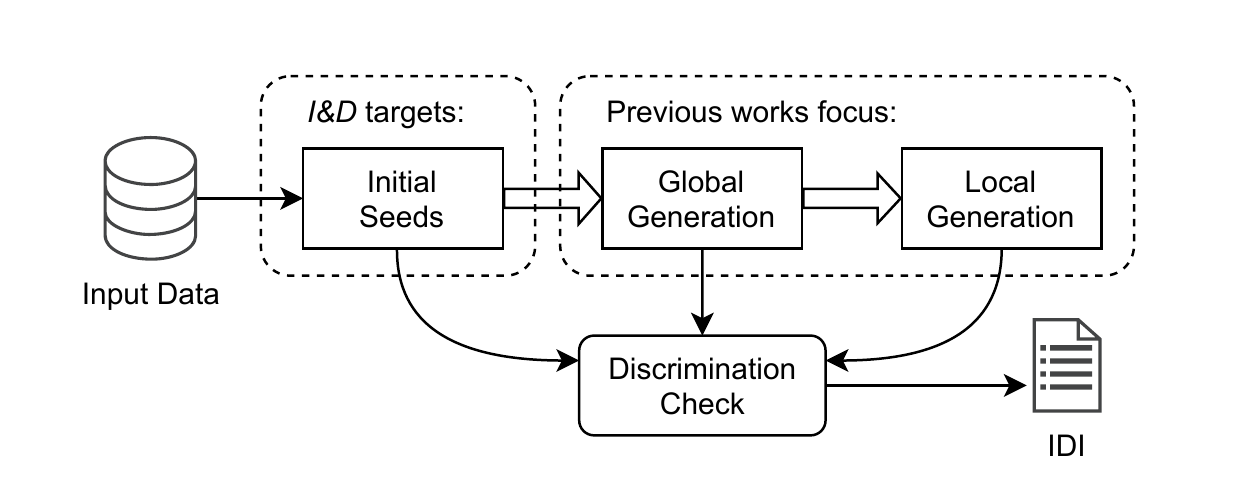}
    \caption{A typical framework for IDI generation. }
    \label{fig:idigap}
\end{figure}

Over the years, several IDI generators have been proposed \cite{udeshi2018automated, aggarwal2019black, zhang2020white, zhang2021efficient}. 
In \autoref{fig:idigap}, we summarize a typical framework for IDI generation.

\point{Initial seeds} 
At first, \textit{initial seeds} are selected from the dataset, which are used in the following steps.

THEMIS \cite{galhotra2017fairness} has been the first work to address individual discrimination testing. 
It chooses initial seeds at random and then tests if they are IDIs without using a global or local generation phase.
AEQUITAS \cite{udeshi2018automated} uses the same random initialization procedure.
SG \cite{aggarwal2019black}, ADF \cite{zhang2020white}, and EIDIG \cite{zhang2021efficient} use K-means to cluster the data. Afterwards, they select initial seeds from each cluster in a round-robin fashion.
The purpose of clustering is to improve the diversity of initial seeds, however, the initial seeds are still randomly sampled, resulting in only few initial IDIs to be chosen. 
To the best of our knowledge, no existing work tackles the problem of improving the quality of IDIs during the initial seeds phase. 
We argue that this phase serves as the foundation for the subsequent global and local generation and is therefore worth to investigate.

\point{Global generation} 
Secondly, the algorithm performs a \textit{global generation} phase to extend from initial seeds in order to cover the input domain $\mathbb{I}$ across a broad range. 

SG creates a decision tree to approximate the machine learning model under test. 
Symbolic execution is used to explore the input domain. 
ADF adopts gradients to maximize the difference between the deep neural networks outputs of two similar instances. 
EIDIG increases the efficiency of ADF through a memorization technique.

\point{Local generation} 
Following the global generation phase, the IDIs found are utilised for \textit{local generation}, which explores their neighborhood for additional IDIs.

AEQUITAS assumes that IDIs are close to one another in a local domain.  
The intuition is to add small modifications to the current IDIs in order to find more IDIs locally.
SG examines locality to evaluate if a small change in the input can influence the model's judgment.  
Depending on the minimal gradient absolute value on each attribute, ADF looks for more discriminatory occurrences among the IDIs' neighbors. 
EIDIG reduces the number of gradient computations performed in ADF by exploiting prior knowledge of gradients.

\subsection{SHAP Value}
To gain a better understanding of machine learning models, SHapley Additive exPlanations (SHAP) \cite{lundberg2017unified} is often adopted. 
SHAP values are calculated as a consistent measure of feature importance, which is also time-saving in terms of computation \cite{frechette2016using}.
The model generates a prediction value for each instance, and a SHAP value is assigned to each feature of the instance. 
Formally, $x_i$ denotes the feature $i$ of an instance $x$. 
The base value $x_{base}$ is the mean of the target class for all instances.
The output of the model prediction result $f(x)$ is:  
$$f(x) = x_{base} + \sum\nolimits_{i=1}^{n}\mathcal{S}(x_i)$$ 
where $\mathcal{S} (x_i)$ is the SHAP value of the $x_i$.
For example, we compute the SHAP value\footnote{https://github.com/slundberg/shap} of the previous example $x$ and $x'$ based on the decision tree model:

$$\mathcal{S}(x): [-0.038, -0.014, 0.012, -0.191, -0.208, -0.260, 0.006, $$
$$-0.001, \textbf{-0.047}, 0.021, 0.005, -0.034, -0.008]$$
$$\mathcal{S}(x'): [0.105, 0.070, 0.107, -0.059, -0.191, -0.122, 0.001, $$
$$-0.001, \textbf{0.181}, 0.036, 0.016, 0.103, -0.003]$$

Note that $x_{base} = 0.757$, indicating the percentage of $y = 0$ predicted labels in the dataset.
In this example, the sum of $\mathcal{S}(x)$ and $x_{base}$ is 0 ($f(x)$), whereas the sum of $\mathcal{S}(x')$ and $x_{base}$ is 1 ($f(x')$).
From the SHAP value, we can determine that gender is the greatest positive feature (0.181) in $\mathcal{S}(x')$.
This implies that it has the strongest positive relationship with the classification outcome of all attributes in the model.
However, gender shows a negative connection (-0.047) with the classification outcome in $\mathcal{S}(x)$.

In this work, the SHAP value is utilized to illustrate the model at the instance level.
Other feature importance metrics, introduced in \autoref{sec:related}, can also be substituted for SHAP to explain the model. 

\section{The I\&D Approach}
\label{sec:design}
\begin{figure*}[]
    \centering
    \includegraphics[width=\linewidth]{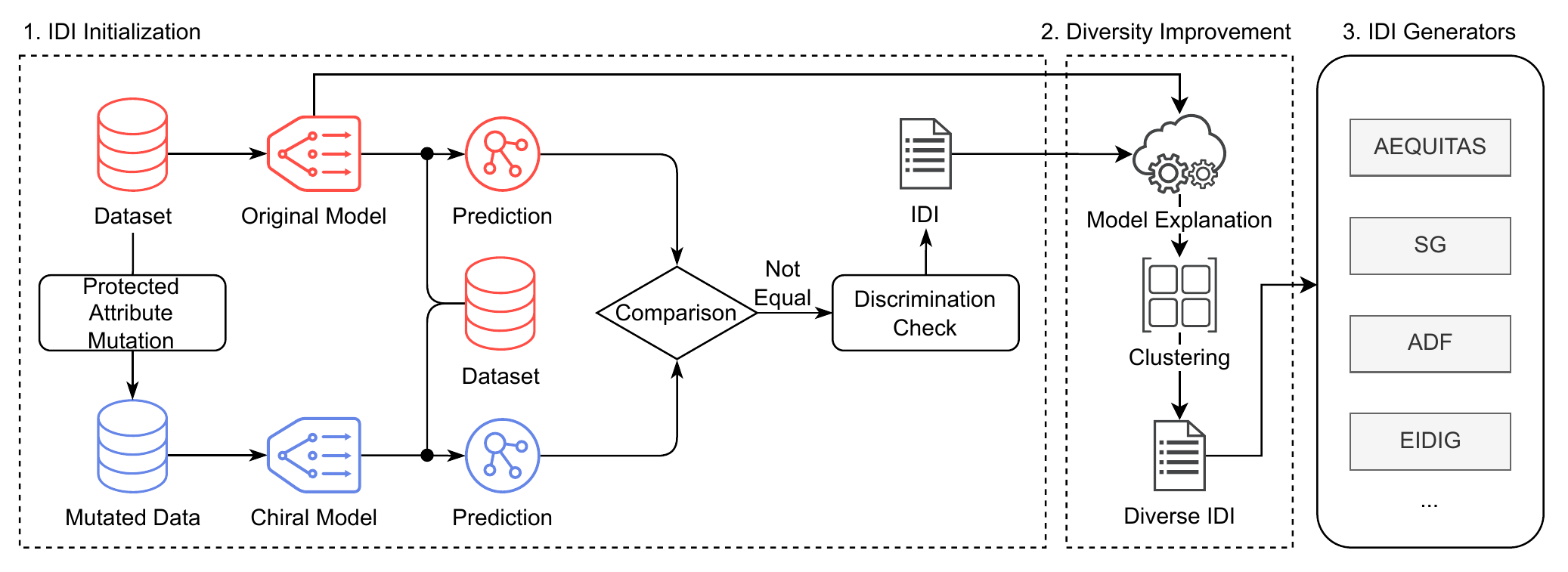}
    \caption{Overview of our proposed approach \name.}
    \label{fig:idi}
\end{figure*}

In this section, we introduce our proposal, dubbed \name, for initial IDI generation.
\autoref{fig:idi} gives an overview of our approach.
First, \name finds initial IDIs through a novel algorithm \textbf{IDI initialization} (\autoref{ssec:init}). 
Then, \name exploits a \textbf{diversity improvement} component (\autoref{ssec:diversity}) to select diverse types of IDIs from those previously found (\autoref{ssec:init}).
Finally, \name it is integrated with the existing state-of-the-art approaches, such as AEQUITAS, SG, ADF, and EIDIG, by simply replacing their initial seed generator (based for example on random sampling or clustering)with the IDIs generated by \name (i.e., instead of using their own initial seeds, we feed each of these state-of-the-art approaches with the initial seeds generated by \name).

\subsection{IDI Initialization}
\label{ssec:init}

To generate initial seeds, we design a novel IDI initialization algorithm based on a ``chiral'' model.
Chirality is a feature of asymmetry that is important in many research areas \cite{cahn1966specification}, such as chemistry, mathematics, and biology.
An object is chiral if it can be distinguished from its mirror image; that is, it cannot be superimposed onto it.
Here we borrow this concept to mutate the protected attributes of a dataset and train a new model, \ie the chiral model, which should yield similar results as the original model. 
The intuition behind using a chiral model is that it has different prediction outputs compared to the original model, and therefore is more likely to find IDIs. 




As shown in \autoref{fig:idi}, the IDI initialization component 
first builds a prediction model, named \textit{original model}.
Then, it mutates the protected attributes in the dataset to obtain \textit{mutated data}.
This is a simple step because we only need to alter the values of protected attributes and leave the rest of the attributes and labels unchanged.
For binary protected attributes, one can simply flip the value from 0 to 1 and from 1 to 0.
For other protected attributes, we modify their value from their input domain at random.
Note that we choose random values rather than a permutation of all values in their input domain, because the mutated data should have the same size as the original dataset. 
Furthermore, because some attributes, such as age, have a broad variety of input domain, supplying every possible value for a protected attribute may lead to a combinatorial explosion. 

After mutating the protected attribute, we utilize the mutated data to train a \textit{chiral model} with the same structure and hyper-parameters as the \textit{original model}.
These two models are then used to predict the label of every instance in the dataset.
If the prediction outputs are different, the instance is subject to a discrimination check.
The discrimination check is based on the definition of IDI, which can identify a set of instances $X$ such that $x$ and any instance $x'$ in $X$ differ only in some protected attributes.
Since \name is not based on a specific machine learning model, \name treats the machine learning model as a black box.

\begin{figure}[t] 
	\centering
	\subfigure[$x$ on original model]{
		\includegraphics[width=\linewidth]{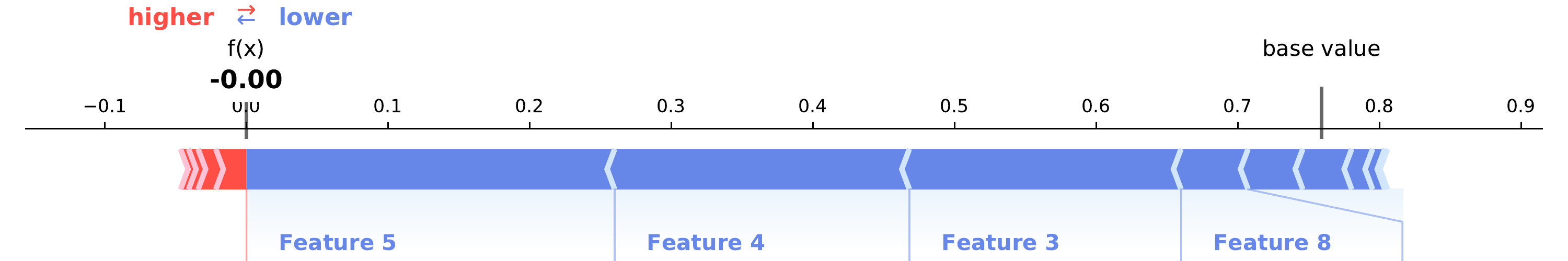}
		\label{fig:ori-u}
	}
	\subfigure[$x'$ on original model]{
		\includegraphics[width=\linewidth]{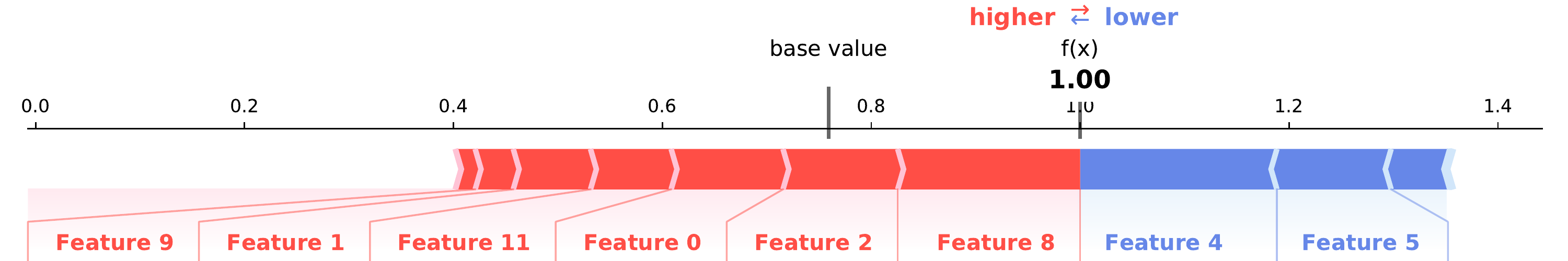}
		\label{fig:ori-n}
	}
	\subfigure[$x$ on chiral model]{
		\includegraphics[width=\linewidth]{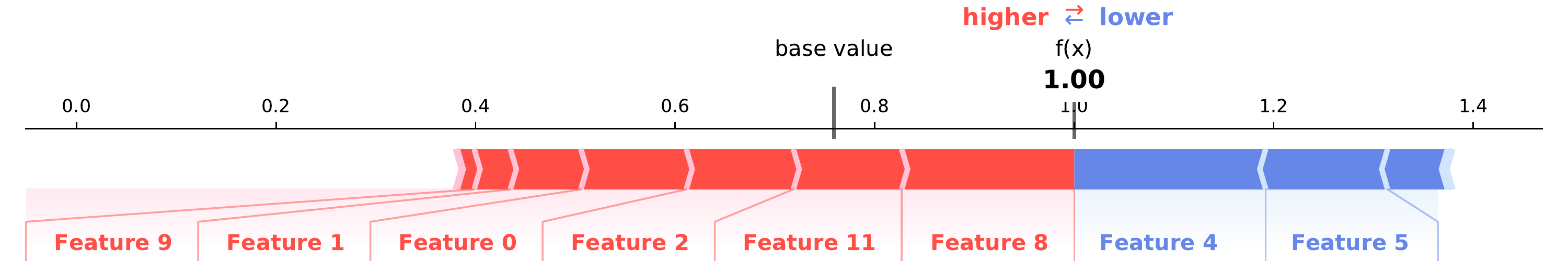}
		\label{fig:chi-u}
    }
	\subfigure[$x'$ on chiral model]{
		\includegraphics[width=\linewidth]{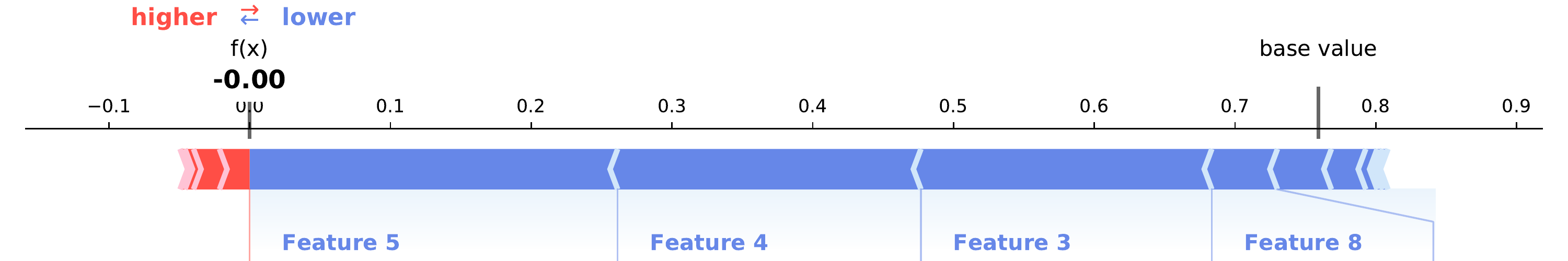}
		\label{fig:chi-n}
	}
	\caption{In the original model, the SHAP value between $x$ and $x'$ is different. The SHAP value of $x$ on the original model (a) is comparable to that of $x'$ on the chiral model (d). (b) and (c) is similar, likewise. It's no surprise that the chiral model on the IDI looks like a mirror image of the original model.}
	\label{fig:case}
\end{figure}

To gain insights into why the chiral model is applicable, we adopt the SHAP value to explain the predictions made by black box models. 
The SHAP value illustrates the contribution of each feature to the final prediction.
We compare the SHAP value of the same instance predicted by the original model and by the chiral model. 
\autoref{fig:case} illustrates a case of the SHAP value on a pair of IDIs.
In this figure, the SHAP value of features is shown beneath the axis.
The red bar illustrates positive values, while the blue bar shows negative values. 
The values are sorted from left to right, with the lowest value being at the left-most position.
The actual prediction output is denoted by $f(x)$, which is the sum of all features' SHAP values.
From \autoref{fig:ori-u} and \autoref{fig:ori-n}, we can observe that $x$ and $x'$ have a different prediction output, since they are a pair of IDIs.
When comparing the chiral model to the original model, the SHAP value of IDIs in the chiral model are almost identical to that of the original model.
For example, \autoref{fig:chi-u} is comparable to \autoref{fig:ori-n}.
As a result, the chiral model's predicted output differs from the original output. 
Note that the SHAP value for $x$ on the chiral model is not the same as $x'$ on the original model. 
The rank of features in \autoref{fig:ori-n} and \autoref{fig:chi-u} differs. 


\begin{figure*}[t]
    \centering
    \includegraphics[width=.8\linewidth]{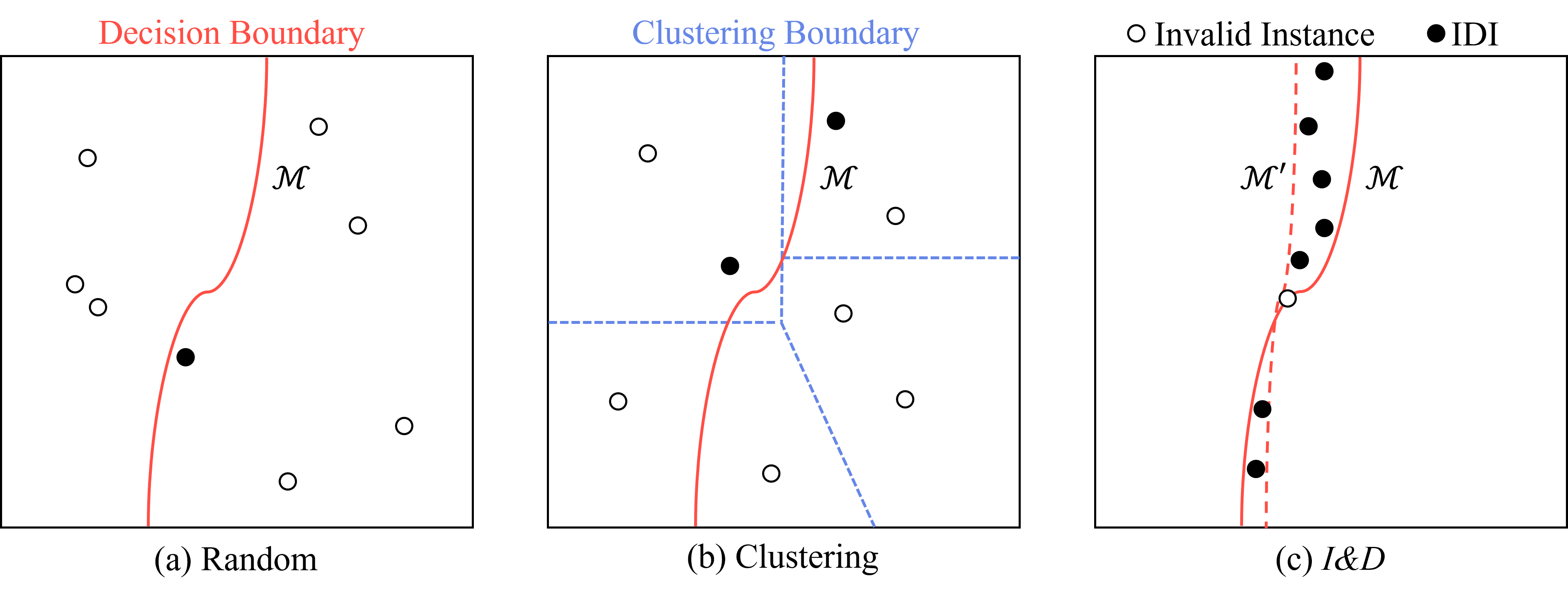}
    \caption{Comparison of different initialization methods. (a) samples the entire input space for 8 instances; (b) divides the input space in four clusters and receives 2 instances from each; (c) illustrates \name, which samples instances close to the decision boundary of the original model (solid) and the chiral model (dashed).}
    \label{fig:toy}
\end{figure*}

\point{Qualitative evaluation}
In the following, we provide the reader with a theoretical explanation of how \name augments and improves existing approaches. This is further supported by the empirical evaluation presented in \autoref{sec:results}.

\autoref{fig:toy} illustrates a sketch map of the three different IDI initialization approaches, namely our proposed \name initialization, Random initialization, and Clustering-based initialization.
Each box in this figure represents the data space in two dimensions.
The red line denotes the decision boundary of a model, while the dashed red line in \autoref{fig:toy} (c) represents the decision boundary of the chiral model. 
For the initial seeds, we chose eight instances, marked by circles.
The black circles are IDIs after the discrimination check, while the white circles are data points without discrimination.
Because IDIs have such a small proportion in the dataset, the random selection approach is obviously difficult to capture.
The clustering-based method divides the data space into numerous groups and obtains seed instances in a round-robin fashion from each cluster.
The purpose of clustering is to increase the diversity of initial IDIs.
Because these cases represent distinct clusters of the dataset, these approaches may gain more IDIs than random sampling.

Unlike existing approaches, \name takes into account both the dataset and the model.
We develop a chiral model, indicated as $\mathcal{M}'$ in this figure, that has a different decision boundary than the original model.
IDIs have a lot of potential for filling in the gap between the original model and the chiral model's decision boundary.
As a result, we can explicitly catch initial IDIs by distinguishing predicted outcomes from the two models.

\subsection{Diversity Improvement}
\label{ssec:diversity}

A desired  property of the initial seeds generation is to obtain a diverse set of IDIs.
In accordance with existing approaches for generating diverse IDIs, we also follow a clustering-based procedure.
However, these approaches group the original dataset directly and this practice might be ineffective because it does not take the prediction model into account. In fact, different models may have different decision-making strategies on the dataset, and the clustering boundary may not precisely overlap with the decision boundary, as shown in \autoref{fig:toy} (b). Therefore, the diversity of IDIs may not be reflected effectively without considering the original model.
On the other hand, it is challenging to directly assess the diversity of IDIs of machine learning models, since these models are often difficult to interpret, besides the testing of such models is desired to be model agnostic, considering them as a black box \cite{aggarwal2019black}.

To tackle these challenges, we design a novel diversity improvement component, which combines the SHAP value and clustering. 
Specifically, we first use the SHAP value to explain the IDIs found in the initialization component, which outputs the SHAP value score of each IDI.
Then, based on their SHAP value, we use DBSCAN \cite{schubert2017dbscan} to cluster these IDIs. 
DBSCAN is a powerful data density-based clustering algorithm that clusters without the need for specifying the target number of clusters \cite{ma2021jump}. 
IDIs in each cluster are then chosen in a round-robin process to obtain IDIs until a specified number of initial seeds is reached.
This upper limit (i.e., search budget) on the number of initial seeds is a parameter adopted by previous studies \cite{agarwal2018automated, zhang2020white, zhang2021efficient}.
If the budget set ends up to be larger than the number of IDIs actually generated, one can fill it up by using random sampling.\footnote{It is worth noting that our experiments (\autoref{sec:results}) show that this had not been needed since \name was always able to identify more initial IDIs than those required by the search budget. In RQ3 (\autoref{ssec:rq3}) we further evaluate the usefulness of this component for fairness testing and improvement.}

We then integrate \name with each existing IDI generation approach (namely AEQUITAS, SG, ADF, and EIDIG) by simply replacing their initial seeds phase with the IDIs generated by \name, which makes our approach very straightforward to integrate with existing IDI generator.

\section{Empirical Study Design}
\label{sec:settings}
In this section, we describe the design  of the empirical study we carried out to assess the effectiveness of our proposal for initial seed generation, dubbed \name.

Specifically, this study aims to address the following research questions (RQs):

\begin{itemize}[]
\item \textbf{RQ1:} How well does \name perform when integrated with existing IDI generation methods?
\item \textbf{RQ2:} How effective is \name for testing different type of machine learning models?
\item \textbf{RQ3:} How do the IDI Initialization and Diversity Improvement components contribute to the success of \name?
\item \textbf{RQ4:} How do the parameters of \name influence its performance?
\end{itemize}

\begin{table}[]
\caption{Datasets used in this study.}
    \vspace{-3pt}
\label{tab:dataset}
\begin{tabular}{lrrll}
\toprule
Dataset   & \# Rows  & \# Attribute & Protected Attribute \\ \midrule
Census & 32561 & 13 & Age, Race, Gender \\
Compas & 10577 & 14 & Age, Race, Gender\\
Bank   & 45211 & 16  & Age  \\
\bottomrule
\end{tabular}
\end{table}

\subsection{Datasets} 
To evaluate the use \name, we employ three commonly used datasets in the software fairness literature~\cite{zhang2020white,zhang2021efficient,hort2021fairea,chakraborty2020fairway,chakraborty2021bias}.
We investigate the following three datasets (with additional information provided in \autoref{tab:dataset}):
\begin{itemize}

\item \point{Adult Census Income (Census)}\footnote{https://archive.ics.uci.edu/ml/datasets/adult}  
Barry Becker retrieved this dataset from the 1994 Census database.
Its objective is to forecast whether a person earns more than \$50,000 per year based on their personal data.

\item \point{Correctional Offender Management Profiling for Alternative Sanctions (Compas)}\footnote{https://github.com/propublica/compas-analysis} 
This is a Broward County criminal history dataset, used to forecast reoffending risks for defendants.

\item \point{Bank Marketing (Bank)}\footnote{https://archive.ics.uci.edu/ml/datasets/bank+marketing} 
This dataset comes from a Portuguese bank and is used to estimate if a customer would sign up for a term deposit based on their information.

\end{itemize}


We pre-process the datasets following the previous work ADF~\cite{zhang2020white} by binning the numerical attributes. 
Also, we remove the feature that directly indicates the prediction label (e.g., risk score in Compas).

\subsection{Machine Learning Models}
Because ADF and EIDIG are only capable of generating IDIs for neural networks, we use fully-connected neural network following the existing studies as the test model in RQ1 (\autoref{ssec:rq1}), RQ3 (\autoref{ssec:rq3}) and RQ4 (\autoref{ssec:rq4}). 
We choose hyperparameters following their settings \cite{zhang2020white, zhang2021efficient}. 
The number of neural network layers was set to 6, the size of neural network hidden states was set to 30, 20, 15, 10, 5, and 1, respectively. 
During training, we used binary cross-entropy loss function, Nadam as the optimizer, 30 training epochs, 128 batch size, and 0.01 learning rate.

For RQ2 (\autoref{ssec:rq2}), we build four widely-used models, \ie Logistic Regression (LR), Support Vector Machine Classifier (SVC), Decision Tree Classifier (DTC), and Multi-layer Perceptron classifier (MLP) following AEQUITAS \cite{agarwal2018automated} from scikit-learn \cite{scikit-learn}. 
The hyperparameters are set to the model's default setting, which we believe has little impact on the results of our experiments because they can attain reasonably high performance (the average F1-score is about 89\%).

\subsection{Implementation}
We implemented \name in Python based on TensorFlow \cite{abadi2016tensorflow} and scikit-learn \cite{scikit-learn}. 
DBSCAN's default $\epsilon$ (distance threshold) is 0.09, and the minimum sample size is 10 \cite{schubert2017dbscan}.
We used the public implementations of four IDI generation approaches, \ie AEQUITAS \cite{udeshi2018automated}, SG \cite{aggarwal2019black}, ADF \cite{zhang2020white} and EIDIG \cite{zhang2021efficient}.
 
We conducted all the experiments on a Linux server with Intel(R) Xeon(R) E5-2640 v4 @ 2.40GHz CPU, 128GB memory, and Ubuntu 18.04 as the operating system.

\subsection{Fairness Testing Metrics}
To assess the potential improvement of our approach to find initial IDIs, we combine \name with four state-of-the-art IDI generation approaches, \ie AEQUITAS, SG, ADF, and EIDIG.
We compare these approaches before and after the integration with \name.
In this study, we consider both effectiveness and usefulness as measures of performance. 
The count of IDIs has been widely-used in previous work \cite{udeshi2018automated, aggarwal2019black, zhang2020white, zhang2021efficient} to evaluate the performance of IDI generation for fairness testing, and thus we adopted the same metric.
Although the count of IDIs is a single measure, it is used in two different manners.

\point{The More the Better for Generation ($\uparrow$)}
Our goal for IDI generation is to effectively generate a large number of IDIs given the computational constraints (``the more IDIs the better'').
We set the limit of initial seeds, and the global and local generation limit both to 100. 
As a consequence, $100,100$ is the maximum number of the two-phase searched instances (100 global and 100*100 local instances).
We also evaluate these limits in RQ4 (\autoref{ssec:rq4}).

\point{The Fewer the Better After Retraining ($\downarrow$)}
Following previous work \cite{zhang2020white, zhang2021efficient}, for each approach, we also used its generated IDIs to retrain the model and then measured the fairness of the retrained model (\ie the number of IDIs for the retrained model) by testing the IDIs generated by the approaches. 
The fewer IDIs remain, the better the IDI generation approach is.

We follow the typical practice to evaluate the classification model accuracy based on the F1-score. 
Each dataset was split, with 60\% of it serving as the training set and 40\% as the test set. 
During the retraining phase, we add IDIs into training data and \changed{keep the parameters of the original model unchanged.} 
The number of epochs is set to 10, because the loss is steady after retraining for 10 epochs.
We experimented with each approach ten times and averaged the results to limit the impact of randomness. 
We observe that the F1-score of the model remains constant before and after retraining, indicating that the model is not overfitting on the IDIs.

Following ADF and EIDIG, we use the notion of majority voting to determine the label of created IDIs based on the decisions of several models, \ie LR, SVC, DTC, and MLP.

\subsection{Threats to Validity}
\label{threats}
We evaluated \name with three datasets. 
They are the most common public benchmarks used in the fairness testing literature.
However, further datasets could be considered for future work to strengthen our results.
Furthermore, the generated IDIs lack ground-truth labels and rely on voting from multi-model prediction outputs  \cite{jiang2020identifying}.
When retraining the model with more produced labels than its initial dataset size, the model may not be helpful.
Therefore, we adopt the Compas dataset following the study of Fairway \cite{chakraborty2020fairway}.
\name is open-sourced and dataset-independent.
If more datasets become accessible in the future, it will be simple to expand our analysis.

The machine learning models used to conduct our experiments are also used in prior studies, such as AEQUITAS \cite{udeshi2018automated} and ADF \cite{zhang2020white}. 
Because the datasets are simple, with a maximum of 16 features, basic models like fully-connected deep neural networks can handle them.
Our approach, on the other hand, is broad and does not rely on any specific models.
Since the main idea of \name is to train a chiral model and compute SHAP value to cluster, which is straightforward to implement even for more sophisticated models.



\section{Empirical Study Results}
\label{sec:results}
In this Section we describe the results of the empirical study as designed in \autoref{sec:settings} to answer our four RQs.

\begin{table*}[t]
\caption{RQ1-Effectiveness: Total number of generated IDIs by each of the existing approach with and without \name ($\uparrow$).}
\label{tab:rq1-1}
\begin{tabular}{cc|rr|rr|rr|rr}
\toprule
Dataset & Protected & AEQ. & AEQ. & SG & SG & ADF & ADF & EIDIG & EIDIG \\ 
 & Attribute & & + I\&D & & + I\&D & & + I\&D & & + I\&D \\ \midrule
\multirow{3}{*}{Census} & Age & 87 & \textbf{4324} & 479 & \textbf{489} & 5055 & \textbf{5716} & 6442 & \textbf{7113} \\
& Race & 23 & \textbf{1500} & 176 & \textbf{249} & 1403 & \textbf{1717} & 2884 & \textbf{4480} \\
& Gender & 57 & \textbf{840} & 134 & \textbf{163} & 781 & \textbf{1134} & 1180 & \textbf{3838} \\ \midrule
\multirow{3}{*}{Compas} & Age & 485 & \textbf{6621} & 630 & \textbf{765} & 6840 & \textbf{7173} & 3324 & \textbf{3995} \\
& Race & 14 & \textbf{4399} & 370 & \textbf{374} & 4521 & \textbf{6722} & 2374 & \textbf{2581} \\
& Gender  & 5 & \textbf{2050} & 182 & \textbf{206} & 2033 & \textbf{2041} & 1295 & \textbf{1424} \\ \midrule
Bank & Age & 189 & \textbf{2839} & 626 & \textbf{662} & 4715 & \textbf{5095} & 4179 & \textbf{6562} \\ \midrule
\multicolumn{2}{c|}{Average} & 123 & \textbf{3225} & 371 & \textbf{415} & 3621 & \textbf{4228} & 3097 & \textbf{4285} \\
\bottomrule
\end{tabular}
\end{table*}

\subsection{RQ1: Performance of \name}
\label{ssec:rq1}

\point{Overall Effectiveness of \name}
The performance comparison between existing IDI generation approaches and their corresponding versions with \name is shown in \autoref{tab:rq1-1}, where bold numbers represent the best results between each pair of versions. 
Each row in this table represents the number of IDIs produced for each dataset and protected attribute.
According to these results, all four approaches with \name outperform the original, existing approaches for all datasets and protected attributes under study.
The average number of generated IDIs across all original approaches is $1,803$, whereas the number created by approaches integrated with \name is $3,038$. This is an improvement of 1.68X.
We also observe that among approaches studied, EIDIG with \name can generate the highest number of IDIs ($4,285$ on average).
Besides, the number of IDIs is improved by 26.2X for AEQUITAS with \name, the largest improvement of all the four approaches. 
This is because AEQUITAS applies random sampling, while the other three approaches apply a clustering method. 

\begin{table}[]
\caption{RQ1-Initialization: IDI initialization rate comparison given 1000 seeds ($\uparrow$).}
\label{tab:rq1-2}
\begin{tabular}{cc|rrr}
\toprule
Dataset  & Protected Attribute & Random & Clustering  & \name \\ \midrule
\multirow{3}{*}{Census} & Age & 3.4\% & 17.1\% & \textbf{78.0\%} \\
& Race   & 2.1\% & 7.4\% & \textbf{40.3\%} \\
& Gender & 0.7\% & 4.9\% & \textbf{35.5\%} \\ \midrule
\multirow{3}{*}{Compas} & Age & 2.3\% & 5.2\% & \textbf{19.8\%} \\
& Race & 9.7\% & 25.8\% & \textbf{31.0\%} \\
& Gender & 7.9\% & 15.3\% & \textbf{44.4\%} \\ \midrule
Bank & Age & 2.9\% & 2.8\% & \textbf{41.5\%} \\ \midrule
\multicolumn{2}{c|}{Average} & 4.1\% & 11.2\% & \textbf{41.5\%}
\\ \midrule \midrule
\multicolumn{2}{c|}{Running Time (s) ($\downarrow$)} & \textbf{0.001} & 0.002 & 2.692 \\
\bottomrule
\end{tabular}
\end{table}

\point{Initialization Comparison}
We further analyze the reason why \name is able to improve the existing IDI generation approaches. 
To this end, we compare \name with the original IDI initialization methods, namely Random Sampling, and Clustering-based Sampling.
We set the limit of initial seeds for each protected attribute in all datasets to $1,000$ for this comparison.
The average number of IDIs is then calculated by repeating each experiment ten times. 

\autoref{tab:rq1-2} shows the IDI initialization rate (i.e., valid IDIs in $1,000$ seeds) comparison among the three approaches, \ie Random, Clustering, and \name.
In terms of IDI initialization rate, we observe that \name surpasses the two compared approaches (\ie Random and Clustering) for all datasets and protected attributes studied herein.  
In particular, the average initialization rate of \name is 41.5\%, while the other two approaches are just 4.1\% and 11.2\%, respectively. 
In addition, the initialization rate for \name ranges from 19.8\% to 78.0\%, suggesting that \name has consistent high initialization effectiveness.  
The clustering-based method takes into account dataset characteristics, whereas the random sampling method does not.
\name considers both the model and the dataset.
In fact, we train a chiral model by mutating the protected attributes of the dataset.
By comparing the original and chiral models, we can explicitly extract individual discrimination.
Therefore, existing approaches that instead do not consider the model do not perform well for IDI initialization. 
\name is the first attempt to overcome this problem, and our findings indicate that it is a promising approach. 

The last row in \autoref{tab:rq1-2} shows the running time of Random, Clustering, and \name. 
Random and Clustering methods take the least amount of time, whereas our approach \name takes 2.692s on average. 
In practice, though, this time is acceptable.
Moreover, the most expensive part of the calculation is training a chiral model which can be done offline.
As a result, with such a low run time, our approach \name is not only effective but also practical.

\begin{table*}[]
\caption{RQ1-Usefulness: Number of remaining IDIs, after retraining with the generated IDIs ($\downarrow$).}
\begin{adjustbox}{max width=\columnwidth}
\label{tab:rq1-3}
\small
\begin{tabular}{cl|rrr|rrr|rrr|rrr}
\toprule
Dataset & Protected & \# Test & AEQ. & AEQ. & \# Test & SG & SG & \# Test & ADF & ADF & \# Test & EIDIG & EIDIG \\ 
 & Attribute & Cases & & + I\&D & Cases & & + I\&D & Cases & & + I\&D & Cases & & + I\&D \\ \midrule
\multirow{3}{*}{Census} & Age & 4413 & 3074 & \textbf{1282} & 968 & 469 & \textbf{385} & 10765 & 4414 & \textbf{3798} & 13554 & 5742 & \textbf{4262} \\
& Race & 1523 & 358 & \textbf{331} & 454 & 27 & \textbf{24} & 3118 & 844 & \textbf{720} & 7359 & 1055 & \textbf{917} \\
& Gender & 850 & \textbf{40} & 112 & 297 & 26 & \textbf{11} & 1915 & 741 & \textbf{223} & 5018 & 979 & \textbf{245} \\ \midrule
\multirow{3}{*}{Compas} & Age & 7106 & 5762 & \textbf{5139} & 1395 & 396 & \textbf{192} & 13995 & 7762 & \textbf{6663} & 7277 & 3093 & \textbf{2836} \\ 
& Race & 4413 & 3734 & \textbf{2141} & 678 & 125 & \textbf{31} & 11228 & 3362 & \textbf{2454} & 4938 & 2715 & \textbf{2189} \\
& Gender & 2055 & 1437 & \textbf{187} & 369 & \textbf{7} & 21 & 4228 & 2543 & \textbf{971} & 2719 & 934 & \textbf{499} \\ \midrule
Bank & Age & 3028 & 1487 & \textbf{553} & 1288 & 245 & \textbf{137} & 9810 & 2531 & \textbf{2367} & 10740 & 4494 & \textbf{2764} \\ \midrule
\multicolumn{2}{c|}{Avg. \# IDI} & 3341 & 2270 & \textbf{1392} & 778 & 185 & \textbf{114} & 7866 & 3171 & \textbf{2457} & 7372 & 2716 & \textbf{1959} \\ \midrule \midrule
\multicolumn{2}{c|}{Avg. F1-score} & -- & 88.1\% & 87.4\% & -- & 88.0\% & 88.1\% & -- & 90.9\% & 91.8\% & -- & 88.9\% & 88.8\% \\ 
\bottomrule
\end{tabular}
\end{adjustbox}
\end{table*}

\point{Usefulness of \name}
\autoref{tab:rq1-3} shows the number of remaining IDIs after retraining the model with the IDIs generated by each of the approaches respectively. 
The column ``\# Test Cases'' indicates the combination of IDIs generated by the existing approach and the approach with \name, which serves as the fairness test cases. 
The models are retrained using the IDIs generated by an approach as well as the approach with \name to allow for comparisons.
For each pair of models, they are tested by the test cases and the number of IDIs remaining are displayed. 
Because the average F1-score (last row in \autoref{tab:rq1-3}) is nearly \changed{identical to the} model before retraining (88.7\%) for all approaches, we can state that the retrained models are not overfitting on the IDIs.
We can also observe that approaches with \name can perform better than the original approaches in most cases.
In particular, the average number of IDIs remained is 2086, while that of approaches with \name is 1481, a reduction of 29\%. 
For each approach with \name, namely AEQUITAS, SG, ADF, and EIDIG, the number of IDIs diminishes by 39\%, 38\%, 23\%, and 28\%, respectively. 

\insight{\textit{Answer to RQ1}: 
\changed{The use of \name improves the average number of IDIs by 1.68X for AEQUITAS, SG, ADF, and EIDIG, as opposed to not using \name.}
Furthermore, after retraining with the IDIs generated by \name, the number of IDIs is decreased by 29\% on average, implying that \name is effective for improving the model's fairness.}

\subsection{RQ2: Testing Different Models}
\label{ssec:rq2}

To test the effectiveness of \name with different machine learning models, we choose AEQUITAS with \name for experiments because SG is not efficient \cite{zhang2020white} and the ADF and EIDIG approaches are both aimed only at neural network models.

\begin{figure*}[t] 
	\centering
	\subfigure[Global Generation ($\uparrow$)]{
		\includegraphics[width=.31\linewidth]{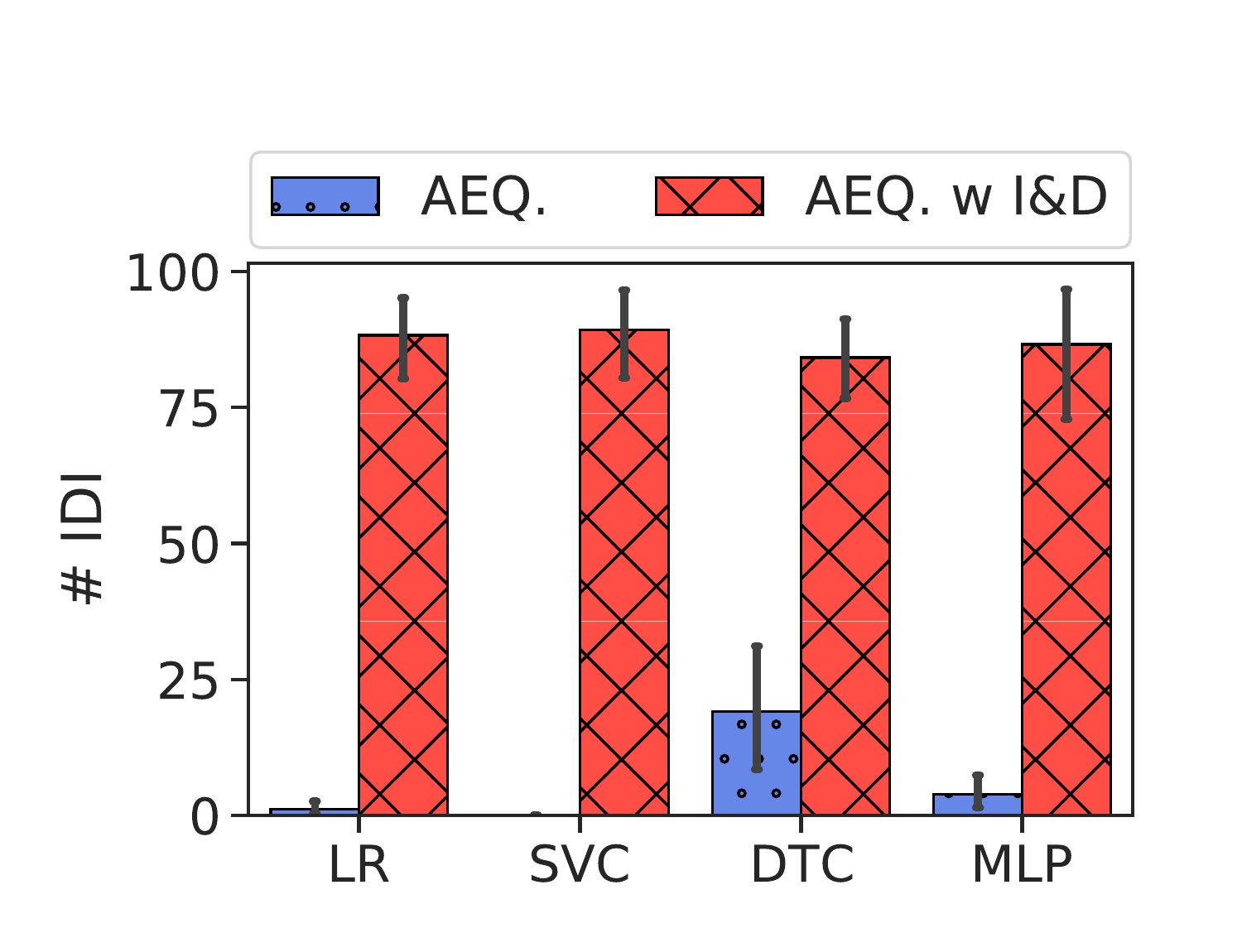}
		\label{fig:global}
	}
	\subfigure[Local Generation ($\uparrow$)]{
		\includegraphics[width=.31\linewidth]{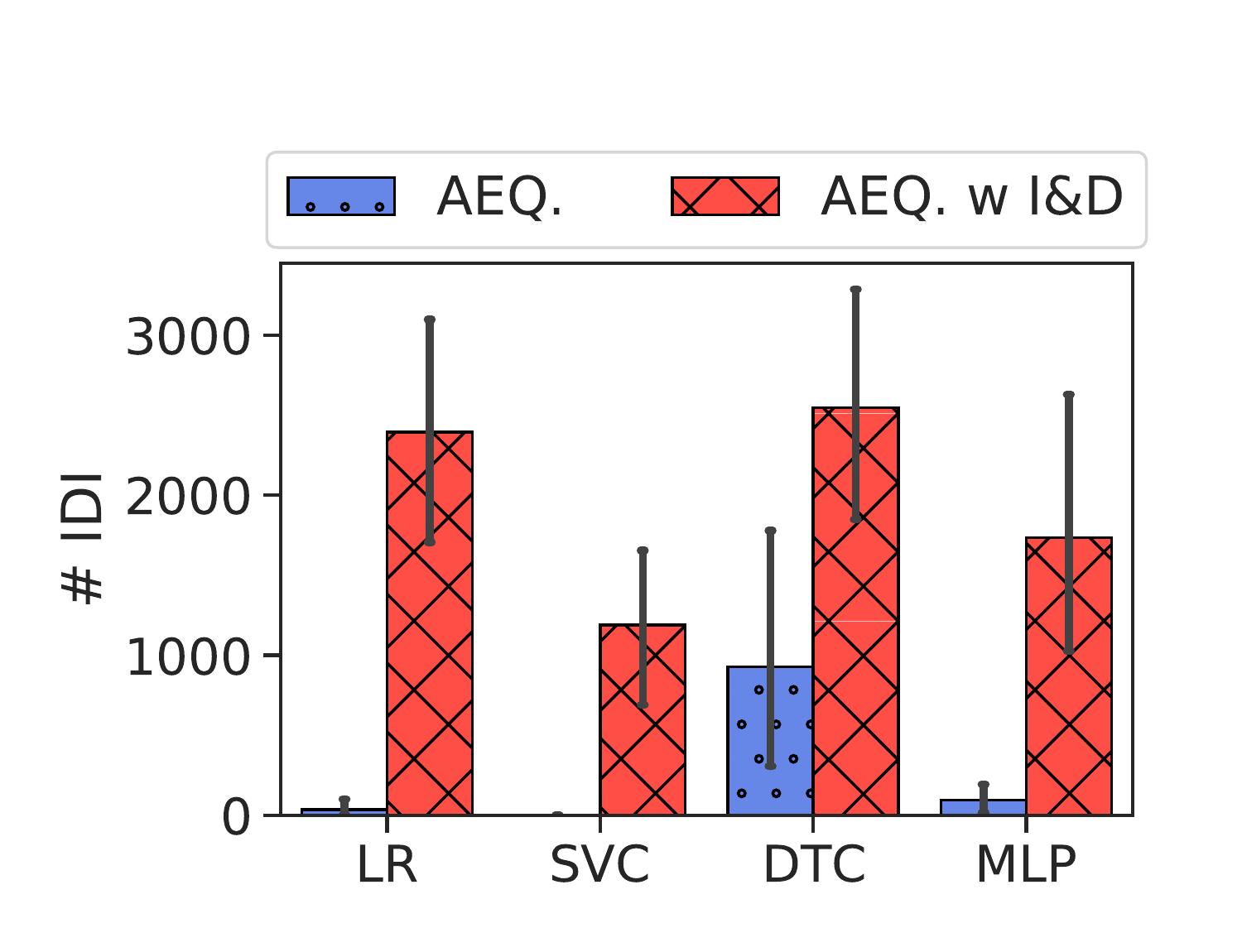}
		\label{fig:local}
	}
	\subfigure[After Retraining ($\downarrow$)]{
		\includegraphics[width=.31\linewidth]{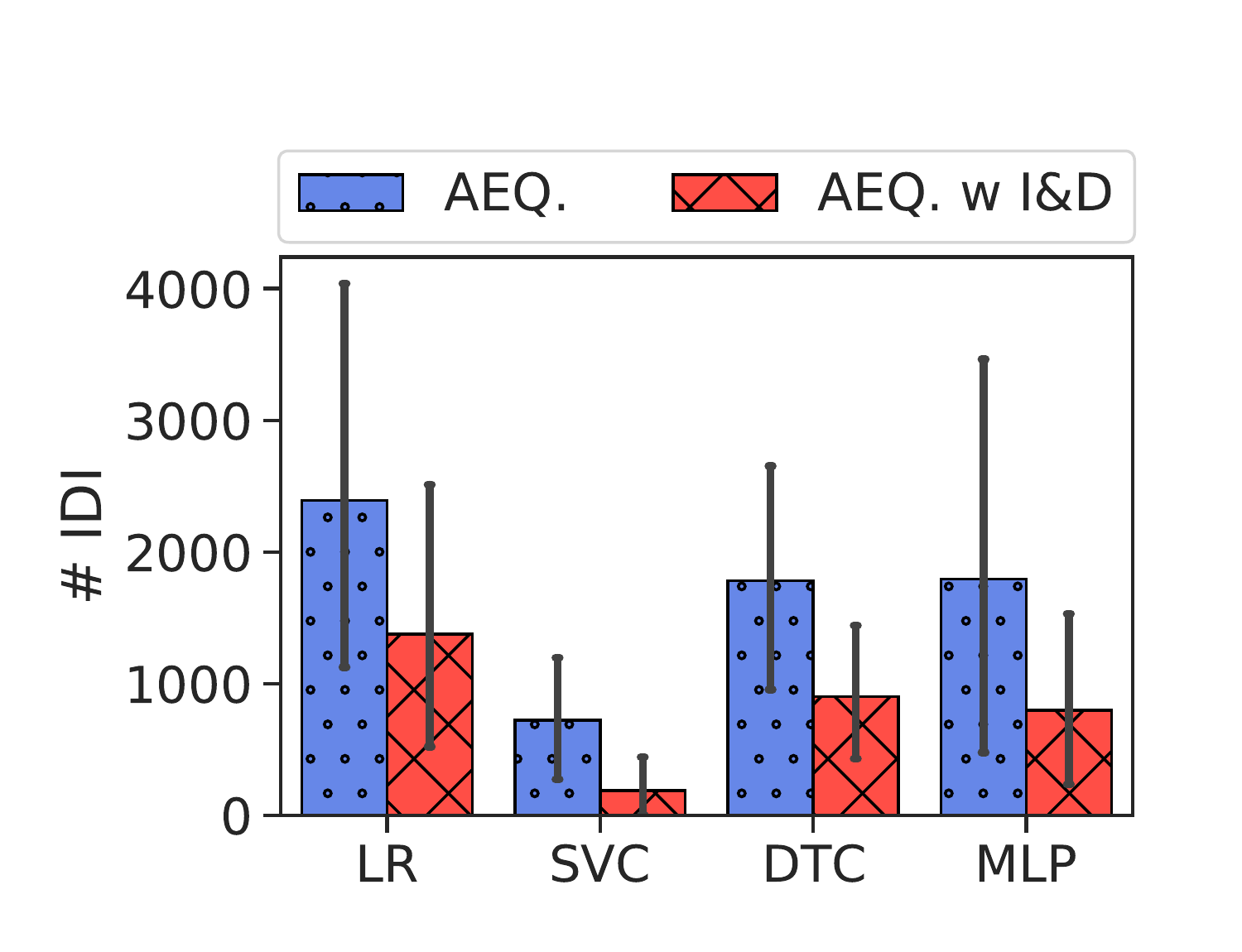}
		\label{fig:retrain}
	}
	\caption{RQ2: Comparison of number of IDIs generated (a, b) and remaining after retraining (c) using different models. \changed{Results are averaged over the three datasets and protected attributes.}}
	\label{fig:rq2}
\end{figure*}

The results of AEQUITAS with or without using \name for different ML models are shown in \autoref{fig:rq2}.
\autoref{fig:global} and \autoref{fig:local} illustrate the global generation and local generation comparison, respectively.
In these figures, we can observe that AEQUITAS with \name beats their original counterparts for all models, namely LR, SVC, DTC, and MLP.
The average number of IDIs generated by AEQUITAS with \name in the global generation phase is 87, whereas the original counterpart is 6, a 14.5X increase.
Note that the initial seeds are used directly in the global generation phase in AEQUITAS, meaning that \name is more effective than random sampling.
In terms of the local generation phase, the average number of IDIs generated by AEQUITAS with \name is $1,967$, whereas the original counterpart is 266, an improvement of 7.4X.
In particular, the LR model has the highest number of IDIs improvement (87 in the global phase and $2,359$ in the local phase on average) of all models, while the SVC model has the smallest number of IDIs improvement (89 in the global phase and $1,190$ in the local phase on average).
The simplest of these four classification techniques is LR, which takes into account the linear relationship between features and the prediction objective.

The number of IDIs remaining after retraining is shown in \autoref{fig:retrain}.
The retraining procedure is similar to that described in \autoref{ssec:rq1}.
According to these figures, the number of IDIs has decreased dramatically from AEQUITAS with \name to the original approaches, regardless of models. 
After retraining using IDIs generated by AEQUITAS, the IDIs remained are 1673, whereas with \name is 816, a 51\% decrease.
The LR model has the greatest number of IDIs drop (1014 on average) of all models, whereas the SVC model has the least number of IDIs decline (534 on average).

\insight{\textit{Answer to RQ2}: 
Regardless of the ML model, \name can successfully help the IDI generator create more IDIs, with an improvement of 14.5X on average.
\name can also help the model reduce discrimination after retraining, with a 51\% reduction on average. }

\begin{table*}[]
\caption{RQ3: Comparison of approaches + I and + I\&D after retraining the model ($\downarrow$).}
\label{tab:rq3}
\small
\begin{tabular}{cl|rrr|rrr|rrr|rrr}
\toprule
Dataset & Protected & \# Test & AEQ. & AEQ. & \# Test & SG & SG & \# Test & ADF & ADF & \# Test & EIDIG & EIDIG \\ 
 & Attribute & Cases & + I & + I\&D & Cases & + I & + I\&D & Cases & + I & + I\&D & Cases & + I & + I\&D \\ \midrule
\multirow{3}{*}{Census} & Age & 7552 & 1711 & \textbf{1631} & 1463 & 288 & \textbf{270} & 16468 &  2614 & \textbf{2255} & 20241 & 2869 & \textbf{2413} \\
& Race & 2515 & 274 & \textbf{127} & 630 & 34 & \textbf{5} & 4847 & 287 & \textbf{177} & 10880 & 1201 & \textbf{508} \\
& Gender & 1259 & 59 & \textbf{42} & 442 & 27 & \textbf{17} & 3072 & 273 & \textbf{238} & 7939 & 186 & \textbf{93} \\ \midrule
\multirow{3}{*}{Compas} & Age & 11482 & 2947 & \textbf{2422} & 2213 & 38 & \textbf{34} & 20915 & 2008 & \textbf{1183} & 10751 & 1648 & \textbf{1227} \\
& Race & 7944 & 5225 & \textbf{5221} & 904 & 515 & \textbf{67} & 16615 & 2610 & \textbf{2008} & 7755 & 1976 & \textbf{1650} \\ 
& Gender  & 3499 & 316 & \textbf{133} & 690 & 113 & \textbf{94} & 6161 & 399 & \textbf{299} & 3788 & 638 & \textbf{465} \\ \midrule
Bank & Age & 5843 & 772 & \textbf{467} & 2004 & 95 & \textbf{83} & 15102 & 1459 & \textbf{1104} & 17004 & 2564 & \textbf{2073} \\ \midrule
\multicolumn{2}{c|}{Average} & 5728 & 1615 & \textbf{1435} & 1192 & 159 & \textbf{87} & 11883 & 1379 & \textbf{1038} & 11194 & 1583 & \textbf{1204} \\
\bottomrule
\end{tabular}
\end{table*}

\subsection{RQ3: Contributions of Components}
\label{ssec:rq3}
To investigate the contribution of the two components in \name, we
compare \name (``\textit{Approach with} \name'') with a modified version that does not take diversity into account (``\textit{Approach with I}'').
To compare the two methods, we count the number of IDIs after retraining the neural network model for these two approaches.

\autoref{tab:rq3} shows the comparison between ``\textit{Approach with I}'' and ``\textit{Approach with} \name'', for AEQUITAS, SG, ADF, and EIDIG.
From these results, we observe that any ``\textit{Approach with} \name'' has fewer IDIs remaining than any ``\textit{Approach with I}''.
More specifically, ``\textit{Approach with} \name'' has 11.1\%, 45.3\%, 24.7\%, and 23.9\% fewer remaining IDIs than the ``\textit{Approach with I}'' on AEQUITAS, SG, ADF and EIDIG, respectively.   
This result indicates \name can improve the model's fairness due to a diverse set of IDIs.

\insight{\textit{Answer to RQ3}: IDIs can be effectively obtained from the target dataset using the IDI initialization component.
The model's fairness can be improved by adding a diversity improvement component.}

\subsection{RQ4: Parameter Sensitivity}
\label{ssec:rq4}

\begin{figure}[t]
    \centering
    \includegraphics[width=.40\linewidth]{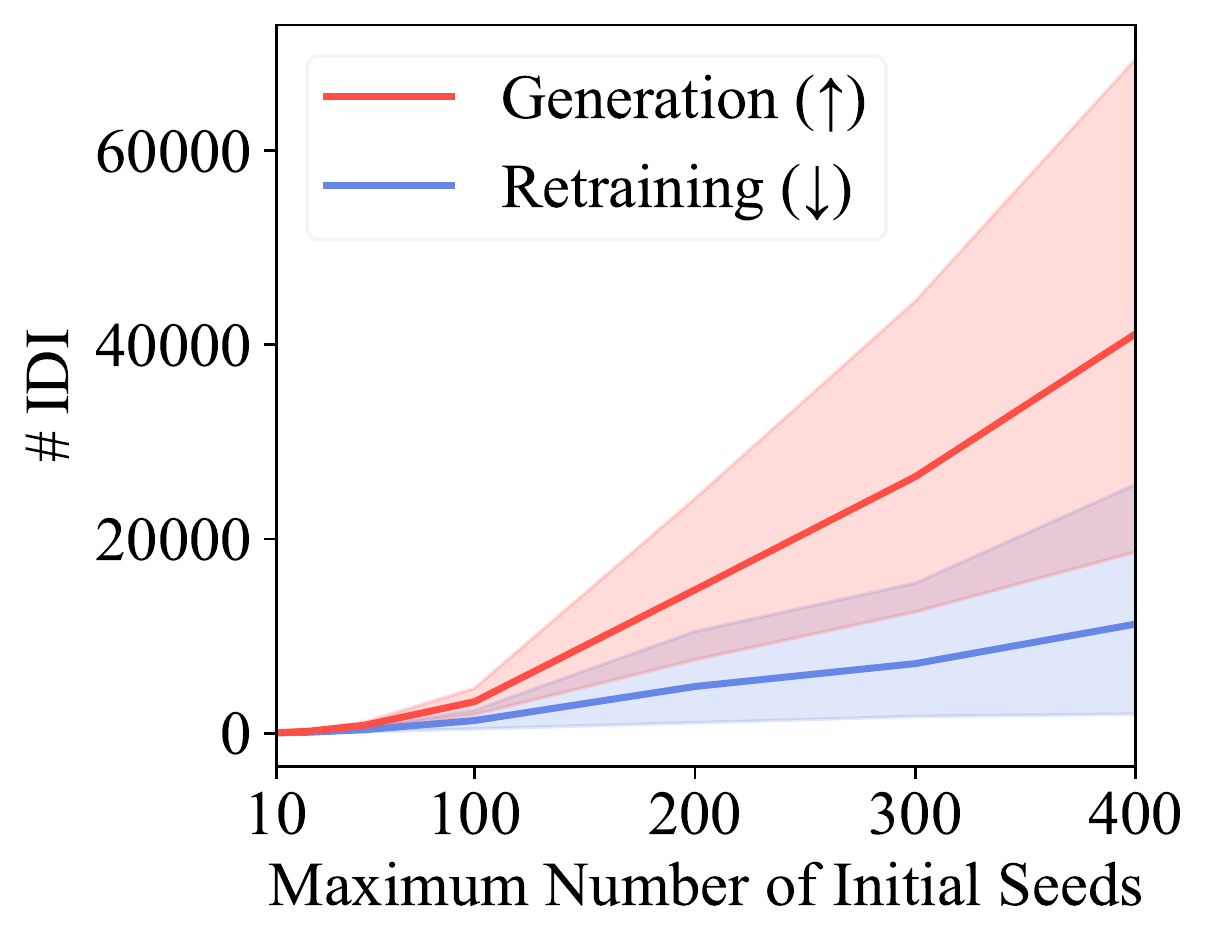}
    
    \caption{RQ4: The average number of IDIs generated and remained after retraining by AEQUITAS with \name under different values for maximum number of initial seeds.} 
    \label{fig:parameter}
\end{figure}

The only parameter of \name is the maximum number of initial seeds. 
Intuitively, the number of generated IDIs depends on the maximum number of seeds for the generating phases, which include initial seeds, global and local generation limits \cite{zhang2021efficient,udeshi2018automated}.
The maximum number of initial seeds should be no greater than the global generation limit, because the global generation uses these seeds as input. 
To keep things simple, we set identical limits for initial seeds, global and local generation.
TO answer RQ4, we analyse what happens when the maximum number of initial seeds ranges from 10 to 400 (see \autoref{fig:parameter}). 
The average number of IDIs generated by AEQUITAS with \name is shown by the solid, red line with error bound.
The dashed, blue line depicts the average number of IDIs remained after retraining.
The number of generated IDIs increases dramatically, because the search space expands exponentially as the limit rises (from 1010 to 160400).
The number of remaining IDIs increases slower than the number of generated IDIs because retraining the model with more IDIs can reduce bias even further.
We do not set a higher limit because the generating time of the AEQUITAS also rises exponentially as the limit increases.
When the limit is 400, it will take more than 14 hours to execute IDI generation on a single protected attribute.
\name, on the other hand, is efficient in handling higher limits. 
Given the limit is 1000, it only takes 2.692 seconds, as shown in \autoref{tab:rq1-3}.

\insight{
\textit{Answer to RQ4}: 
The maximum number of initial seeds, as well as global and local generation limits, influence the final number of generated IDIs.
As the limits increase, the number of generated IDIs and generation time increase dramatically, whereas the number of remained IDIs after retraining increase slower.
}

\section{Related work}
\label{sec:related}

\subsection{Fairness in Software Engineering} 
Fairness is a critical non-functional testing property of data-driven applications and machine learning software~\cite{zhang2020machine}.
As such, it has received an increasing attention from both the software engineering~\cite{zhang2021ignorance,brun2018software,chakraborty2020fairway,ase2021nier} and machine learning research communities~\cite{berk2017convex,ai360}.
Among others, Brun \etal~\cite{brun2018software} named this ``software fairness'' and called for software engineers to combat such discrimination and build fair software.
Since then, fairness concerns have been addressed in different stages of the software development process~\cite{soremekun2022software}.
In addition to software testing~\cite{chen2022fairness}, fairness with regards requirements~\cite{finkelstein2009search} and the design of fair algorithms~\cite{chakraborty2020fairway} have been investigated.
Fair design approaches have been introduced in various stages of the development process, including pre-processing, in-processing, and post-processing~\cite{mehrabi2021survey, chakraborty2021bias, biswas2021fair, zhang2021ignorance}. 
Fairkit \cite{johnson2020fairkit,hort2022bias},
AI Fairness 360 \cite{ai360}, and Fairea \cite{hort2021fairea} aim to mitigate bias.
Our study complements these methods for detecting and mitigating bias.


\subsection{Individual Discriminatory Testing}
Various types of approaches have been proposed for fairness testing of machine learning models in the past few years.
THEMIS \cite{angell2018themis, galhotra2017fairness} first defined software fairness testing in terms of individual discrimination. 
However, THEMIS is inefficient because it relies on random sampling without generating IDIs.
AEQUITAS \cite{udeshi2018automated}, SG \cite{aggarwal2019black}, ADF \cite{zhang2020white} and EIDIG \cite{zhang2021efficient} are a series of IDI generation frameworks. 
A detailed description of these approaches is presented in \autoref{sec:bg_individual}. 
They concentrate on generating IDIs effectively and efficiently in the global and local generation phases but overlook the importance of the seed selection phase.
\changed{To fill this gap, in this work we have proposed a novel way to generate initial seeds, which is able to further improve IDIs effectiveness and diversity. Our work is orthogonal to previous work, as it can be applied to existing IDIs generator by simply replacing their own initial seed generation strategy.}

\subsection{Model Explanation}
In addition to SHAP~\cite{lundberg2017unified} there exist other model explanation techniques.
For example, Local Interpretable Model-agnostic Explanation (LIME) illustrates the model with a decision tree-like structure.
Decision rules \cite{agrawal1994fast} are approaches that are easily understood by humans.
However, they are only useful when they have human-reasonable size. 
In this work, we require an explanation approach at instance level to calculate feature importance \cite{zien2009feature, guidotti2018survey}.
The Shapley value \cite{roth1988shapley} is the foundation for various approaches that credit a machine learning model's prediction on an instance to its underlying features.
The Shapley value can be calculated using a variety of algorithms \cite{sundararajan2020many}, from which we choose the SHAP value \cite{lundberg2017unified}, since it is capable of efficiently explaining a wide range of models.  

\section{Conclusion}
\label{sec:conclusion}

Fairness testing can be used to detect individual discriminating instances (IDIs) and asses an AI system's fairness.
In this paper, we have proposed a novel initialization approach for fairness testing, \name, to aid in the initial phase of IDI generation.
\name compares the prediction output between the original model and a chiral model and uses the SHAP value to improve the diversity of IDIs.
The usefulness of \name was demonstrated through an empirical study on three widely-used datasets for fairness testing research. The average number of IDIs generated  by using our \name approach achieves improvements of 1.68X and exceeds that of the existing approaches.
Furthermore, we discover that by utilizing the generated IDIs to retrain the model and test IDIs again, the remaining IDIs are reduced by 29\%, thus outperforming other approaches.
We also show how the fairness of widely-used models like Logistic Regression, Support Vector Machines, Decision Trees, and Neural Networks can be improved by using \name. 
The contributions of the key components are also supported by our experiments.
\changed{Overall, the results show that the initial seed phase is an important step in the fairness testing procedure, for increasing the number of generated IDIs and proceeding fairness improvements, and should receive more attention.
In future work, we aim to investigate additional methods for selecting initial seeds for comparisons with the chiral model.}


\begin{acks}
We would like to thank the discussion and strong support from Junjie Chen and Chuan Luo.
Max Hort and Federica Sarro are supported by the ERC Advanced fellowship grant no. 741278 (EPIC). 
Hongyu Zhang is supported by the Australian Research Council (ARC) Discovery Projects (DP200102940, DP220103044)
\end{acks}

\bibliographystyle{plain}
\bibliography{references}

\end{document}